\documentclass[%
preprint,
 amsmath,amssymb,
 aps,
superscriptaddress]{revtex4-2}

\usepackage{graphicx}
\usepackage{dcolumn}
\usepackage{bm}
\usepackage[alsoload=synchem]{siunitx}
\usepackage[colorinlistoftodos]{todonotes}

\begin{document}


\title{Flow network controlled shape transformation of a thin membrane through differential fluid storage and surface expansion}

\author{Yongtian Luo}
\affiliation{Department of Physics and Astronomy, University of Pennsylvania, Philadelphia, Pennsylvania 19104, USA%
}%
\author{Che-Ling Ho}
\affiliation{Department of Biology, University of Pennsylvania, Philadelphia, Pennsylvania 19104, USA%
}%
\author{Brent R. Helliker}
\affiliation{Department of Biology, University of Pennsylvania, Philadelphia, Pennsylvania 19104, USA%
}%
\author{Eleni Katifori}%
\affiliation{Department of Physics and Astronomy, University of Pennsylvania, Philadelphia, Pennsylvania 19104, USA%
}%
\newcommand{\dif}{\mathrm{d}}
\newcommand{\vek}[1]{\boldsymbol{#1}} 





\begin{abstract}
The mechanical properties of a thin, planar material, perfused by an embedded flow network, can be changed locally and globally by the fluid transport and storage, resulting in small or large-scale deformation, such as out-of-plane buckling. Fluid absorption and storage eventually cause the material to  locally swell. Different parts can hydrate and swell unevenly, prompting a differential expansion of the surface. In order to computationally study the hydraulically induced differential swelling and buckling of such a membrane, we develop a network model that describes both the membrane shape and fluid movement, coupling mechanics with hydrodynamics. We simulate the time-dependent fluid distribution in the flow network based on a spatially explicit resistor network model with local fluid-storage capacitance. The shape of the surface is modeled by a spring network produced by a tethered mesh discretization, in which local bond rest lengths are adjusted instantaneously according to associated local fluid content in the capacitors in a quasi-static way. We investigate the effects of various designs of the flow network, including overall hydraulic traits (resistance and capacitance) and hierarchical architecture (arrangement of major and minor veins), on the specific dynamics of membrane shape transformation. To quantify these effects, we explore the correlation between local Gaussian curvature and relative stored fluid content in each hierarchy by using linear regression, which reveals that stronger correlations could be induced by less densely connected major veins. This flow-controlled mechanism of shape transformation was inspired by the blooming of flowers through the unfolding of petals. It can potentially offer insights for other reversible motions observed in plants induced by differential turgor and water transport through the xylem vessels, as well as engineering applications. 
\end{abstract}

\maketitle

\newpage

\section{\label{sec:intro} Introduction}

Natural shape-morphing systems, which are ubiquitous in various living organisms and are of great interest for soft matter and biological physics, have been extensively studied by both theorists and experimentalists, in attempts to understand and reproduce their dynamic morphological properties in biomimetic materials. Shape-morphing phenomena which depend on the hydraulics and mechanics of liquid flowing through living matter are widespread in animals, plants and fungi, accommodating diverse needs for speeds and length scales of motions. The ability to absorb or release fluids in the system and the resulting variations of cell hydrostatic pressure (turgor) play a critical role in most common mechanisms for the generation and control of these shape transformations. For instance, in many invertebrate animals, from nematodes (without blood vessels) to mollusks (with blood vasculature), hydrostatic skeletons have evolved and are used to harness turgor pressure, maintain structural rigidity and regulate body movements \cite{Monahan-Earley2013}. Likewise, plants utilize turgor pressure to generate and control both irreversible (such as growth) and reversible motion and deformation in a wide range of time scales by employing several distinct mechanisms \cite{Dumais2012, Forterre2013}. For example, the Venus flytrap achieves a swift snapping closure by a mechanical instability involving hydrostatic accumulation. This is in stark contrast with other slow, gradual movements driven purely by water transport whose speed is limited by diffusion \cite{Forterre2005, Skotheim2005}. The latter hydraulically driven transformations include the folding and shrinking of pollen grains which dehydrate in a dry environment \cite{Katifori2010a, Couturier2013}, and the swelling and expansion of thin and nearly flat plant structures like leaves and flower petals, which can be facilitated by fluid flow networks to overcome the spatiotemporal limitations of water diffusion on transport efficiency \cite{Katifori2018}. 

\begin{figure}[hbt!]
\centering
\includegraphics[width=\textwidth]{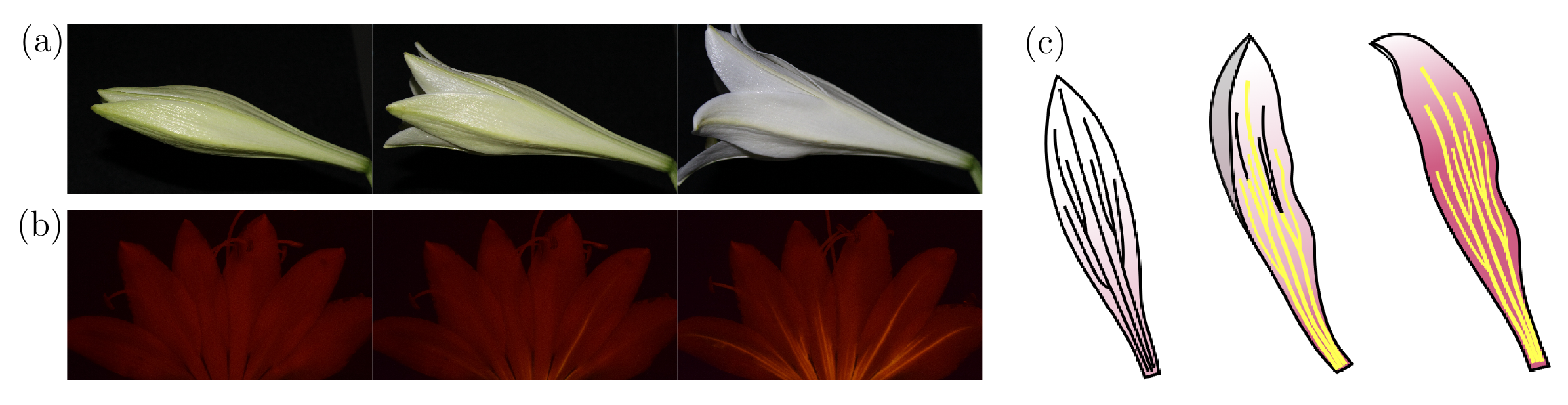}
\caption{Biological inspiration of flow-controlled shape transformation. In each subfigure, the changes proceed from left to right. (a) Flower blooming of Easter lily (\emph{Lilium longiflorum}), snapshots taken by a Canon EOS 1200D camera. (b) The flow of fluorescence dye in the xylem venation of plantain lily (\emph{Hosta sp.}) petals in a flower that has already bloomed, opened for visualization. Images captured by a Nikon D3300 camera. See Appendix for the experimental approach to create these fluorescent images. (c) Cartoon illustration of the effect of flow network on petal deformation. As incoming liquid flux (yellow) fills up empty vessels (black), the originally flat surface expands differentially and buckles out of the plane.}
\label{inspire}
\end{figure}

Examples of such flow-controlled deformations of thin sheets are manifested in the motion of petal expansion in reversible flower blooming, which is a biologically important phenomenon where petals open and close repeatedly in a 24 hour cycle. (An illustration of flower blooming is shown in Figure \ref{inspire} (a).) Reversible flower opening is thought to have evolved to take place when plants need to attract insect pollinators that are only present during a specific time window in a day (e.g.\ during the night) \cite{VanDoorn2003, VanDoorn2014}. In general, large-scale petal deformations often take place during the flowering process (reversible or irreversible), with time scales ranging between minutes and days. They frequently result from differential tissue growth and cell elongation in the organ, which requires the regulation of water flow and turgor pressure \cite{Beauzamy2014}. In a hydration process, cells of different petal segments absorb water and become saturated at different rates, and swell under turgor unequally as a result of the uneven distribution of instant local water content, causing a differential expansion of petal surface which may go through remarkable buckling and shape morphing. The hydraulic behaviors of petals can partially depend on the embedded vascular networks for fluid delivery (illustrated in Figure \ref{inspire} (b)), and are in many ways similar to leaf hydraulics with the same driving force of water potential \cite{Sack2006}. Petal venation systems consist of both phloem and xylem vessels which can form a rich diversity of network architectures including various hierarchies, though usually accompanied by far lower density of stomata and much less transpiration on the surface than leaves \cite{Roddy2013, Zhang2017}. Petals have also been found to make use of large hydraulic resistance (reduced conductance) and capacitance (for water storage) to sustain floral water status and turgor pressure, strengthening flower structural rigidity in a similar fashion to animal hydrostatic skeletons \cite{Roddy2016, Roddy2019}. The considerations of the presence of fluid transport vasculature, the absence of noticeable water loss through evaporation, and the existence of fluid-storage capacitance are all crucial for identifying the specific dynamics of turgor change and subsequent petal motion and deformation. The biological inspiration for this work is summarized in Figure \ref{inspire}, in which the cartoon in (c) depicts the effect of liquid movement through flow network on the surface expansion and shape morphing (buckling) of a single petal.

In this work, we focus on the role that flow networks with local fluid-storage function play in controlling shape transformations of thin sheets actuated by fluid flows. We are particularly interested in differential surface expansion as inspired by flower petals, and implement a simplistic theoretical model that couples hydraulic networks to deformation mechanics. Specifically, the model numerically simulates a variety of time-dependent swelling and deformation pathways, controlled by multiple uniform or hierarchical designs of flow network architecture and influenced by hydraulic traits. This way we quantitatively explore the extent to which the venation structure can affect the emergence of deformed shapes and guide these pathways, while ignoring any transpiration. Mechanical inhomogeneities in the tissue (in particular multiple layers that swell differentially), spatiotemporal variations of fluid storage ability, details of liquid transport mechanism and biochemical regulating factors in real-life petals \cite{VanDoorn2014, Beauzamy2014} are all beyond the scope of our minimal physical model. 

Conventionally, studies of petal differential expansion have concentrated on the growth of certain parts of a petal, like the inner layer, midrib, edge \cite{Liang2011}, or epidermal cells versus underlying layers \cite{Huang2017}. Motivated by these findings, engineers have created flower-like biomimetic thin materials that exhibit large-scale deformations by incorporating anisotropic growth moduli in the plates \cite{Gladman2016, VanRees2017}. The mechanical coupling between hydraulics and elasticity can play a role in deforming such a thin film with a flower shape through capillary forces \cite{Roman2010}. We aim to draw attention to an alternative mechanism that can actuate large-scale deformations, i.e.\ fluid flows through venation networks, which can substantially alter the differential stored fluid distribution and subsequent surface expansion patterns. We present our modeling methods and results in Sections \ref{sec:mm} and \ref{sec:results}, respectively, in which the hierarchical network designs are inspired by leaf venation models with either branching or reticulate (loopy) structure \cite{Katifori2010, McKown2010}, as petal venation is visually similar to the variety of hydraulic hierarchies contained in both monocotyledonous and dicotyledonous leaves (see micrographs in References \onlinecite{Zhang2017} and \onlinecite{Sack2013}.) In Section \ref{sec:disc}, we discuss our results and their implications for both plant biology and biomimetics.

\section{\label{sec:mm} Simulation methods for the coupled flow and spring network model}

\begin{figure}[hbt!]
\centering
\includegraphics[width=0.6\textwidth]{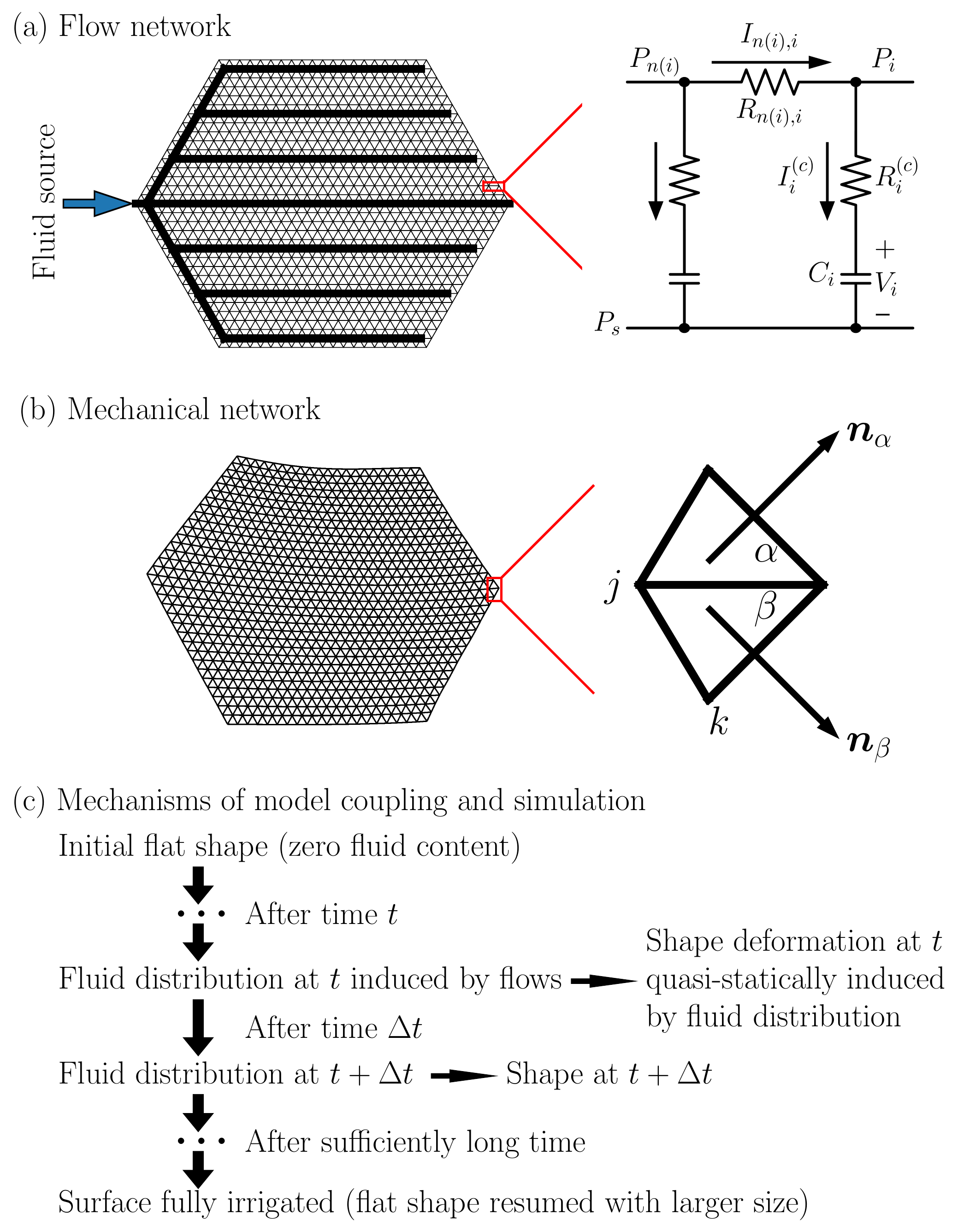}
\caption{The hydrodynamic-mechanical coupling model, consisting of a fluid-storing flow network (a) and an overlapping mechanical network (b), which models the shape changes. In (a), the same hierarchy composed of major (thicker) and minor veins is used to induce the shapes in Figure \ref{fork} (b). The specific hydraulic connectivity for the resistors and capacitors is shown for two neighboring nodes. In (b), examples of nodes ($j$ and $k$), bonds, and faces ($\alpha$ and $\beta$), as well as their normals ($\vek{n}_a$ and $\vek{n}_b$) are shown. Specific coupling methods and simulation procedures are given in (c).}
\label{network}
\end{figure}

The hydrodynamic-mechanical coupling model, which we develop to simulate the spatiotemporal dynamics of fluid transport and storage (using a flow network) and to study the time-dependent resulting shape change (using a mechanical network), is elaborated in Figure \ref{network}. The two network systems comprising the model, though essentially different in their physical properties and functions, have identical topology and connections, overlapping in real space with one-to-one correspondence between their edges and nodes. The flow network operates independently from the mechanical one, and changes in stored fluid are assumed to instantly affect the properties of the mechanical network and induce a deformation as detailed in Figure \ref{network} (c). While the state of the flow network affects the state of the mechanical one, to first-order approximation we assume that their relationship is uni-directional, i.e.\ that the flow network is not affected by the deformation of the mechanical one. In this section we describe in detail the properties and function of each network, and specify the coupling and interactions between the flow and mechanical networks.

\subsection{\label{sec:flow} Numerical simulation of the flow network}

We numerically simulate the dynamics of the flow network by making use of a spatially explicit capacitive model that we developed previously for modeling leaf hydraulics with water-storage functions \cite{Luo2021}. Figure \ref{network} (a) illustrates an example of such a network model consisting of a venation system in which fluids flow (such as xylem vessels), where major veins with smaller resistivity (and larger conductivity) than the rest of the network are represented by thicker lines. (The simulation results of the same network can be found in Figure \ref{fork} (b).) Electrical circuit (ohmic) analogues of hydraulic traits, which include both abilities to allow fluid flow (measured by conductance or resistance, the inverse of conductance) and to store fluid (measured by capacitance), are also shown for a vein segment (representative of every vein segment) between a node with index $i$ and one of its neighbors with index $n(i)$. The driving force of the fluid flow, pressures $P_i$ and $P_{n(i)}$ which are analogous to electric potentials, are labeled at the respective nodes. The direction of the flow current $I_{n(i),i}$ between them, analogous to electric current, is assumed to point from node $n(i)$ to $i$ over a hydraulic resistance $R_{n(i),i}$ defined on the vein segment. In a hierarchical network like the one in the figure, major veins are characterized by smaller hydraulic resistance than minor veins. To model the function of fluid storage (reservoir) at each node, we define a baseline pressure $P_s$ which is uniform throughout all the nodes. In a water-storing plant leaf or petal, the pressure $P_i$ at a node corresponds to xylem water potential and $P_s$ is controlled by a baseline osmotic potential. This baseline potential is determined by the most negative osmotic potential of reservoir cells (which contain aqueous solutes) when the cells hold a minimum water amount and can still behave like linear capacitors \cite{Smith1987, Jones2013, Luo2021}. In the electric analogue, a current $I_i^{(c)}$ fills the storage capacitor $C_i$ with voltage drop $V_i$ across its terminals (whose physical meaning in a leaf or petal is the hydrostatic pressure or turgor of storage cells) with polarity shown in Figure \ref{network} (a) by the plus and minus signs, through a hydraulic resistance $R_i^{(c)}$. The capacitor is shown in the process of being charged, increasing in fluid content which can be calculated readily as $W_i=C_i V_i$ anytime throughout the process. Unlike a plant leaf, the material we consider here does not contain distributed fluid sinks, i.e.\ surface pores (such as stomata) through which the fluid can evaporate and transpire into the atmosphere. Therefore all fluids entering the network will eventually be absorbed and stored in capacitors when the system approaches a steady state, in which the surface is fully hydrated and pressure $P_i$ of all the nodes is uniform and equal to the pressure of the fluid source.

Our simulation methods follow to a large degree the methods described in Reference \onlinecite{Luo2021}, adapted to the modeling in this work as follows. We denote $n(i)$ to be the label of any neighboring node of $i$, and then from mass conservation, we have $\sum_{n(i)}I_{n(i),i}=I_i^{(c)}$. We consider $I_{n(i),i}=(P_{n(i)}-P_i)/R_{n(i),i}$ and $I_i^{(c)}=\partial[C_i(P_i-P_s-R_i^{(c)}I_i^{(c)})]/\partial t=C_i\partial P_i/\partial t-C_i R_i^{(c)}\partial I_i^{(c)}/\partial t$ (as all hydraulic traits, as well as $P_s$, are time-independent). Combining these equations, we eventually have:
\begin{equation}
\Big(\sum_{n(i)}\frac{1}{R_{n(i),i}}+\frac{1}{R_i^{(c)}}\Big)\frac{\partial P_i}{\partial t}-\sum_{n(i)}\frac{1}{R_{n(i),i}}\frac{\partial P_{n(i)}}{\partial t}=\frac{1}{C_i R_i^{(c)}}\sum_{n(i)}\frac{P_{n(i)}-P_i}{R_{n(i),i}}.
\end{equation}
The set of equations for all nodes $i=1,2,\dots,N$ where $N$ is the total number of nodes, is further organized into a matrix equation $\vek{A}\vek{x}=\vek{b}$, where the vector to be solved is:
\begin{equation}
\vek{x}=\left(\frac{\partial P_1}{\partial t},\frac{\partial P_2}{\partial t},\dots,\frac{\partial P_i}{\partial t},\dots,\frac{\partial P_N}{\partial t}\right)^{\text{T}}
\end{equation}
with $\text{T}$ indicating the transpose. At time $t$, the $i$th element of vector $\vek{b}$ is:
\begin{equation}
\vek{b}_i=\frac{1}{C_i R_i^{(c)}}\sum_{n(i)}\frac{P_{n(i)}(t)-P_i(t)}{R_{n(i),i}}
\end{equation}
and the elements in the invertible and symmetric matrix $\vek{A}$ are:
\begin{equation}
\vek{A}_{i,j}=\left\{\begin{array}{lr}\sum_{n(i)}1/R_{n(i),i}+1/R_i^{(c)}&i=j\\ -1/R_{j,i}&j\text{ is neighbor of }i\\ 0&i\neq j\text{ \& }j\text{ is not neighbor of }i\end{array}\right..
\end{equation}
For a node $i$ connected to a fluid source pressure $P_p$, we have $\vek{b}_i=1/(C_i R_i^{(c)})[(P_p-P_i)/R_{p,i}+\sum_{n(i)}(P_{n(i)}-P_i)/R_{n(i),i}]$ and $\vek{A}_{i,i}=1/R_{p,i}+\sum_{n(i)}1/R_{n(i),i}+1/R_i^{(c)}$, where $R_{p,i}$ is the resistance between $i$ and the external node of fluid source.

We start the numerical simulation from an initial state $P_i(t=0)=0$ for the whole network. We define the baseline pressure $P_s=0$, and thus have fluid content $W_i(t=0)=0$ for all the nodes. A constant positive source pressure $P_p$ is connected to a node (here the leftmost node of the midline) at $t=0$ to initiate the dynamic changes of $P_i$ and $W_i$. At each simulation time $t$, we calculate $\vek{b}$ and then $\vek{x}=\vek{A}^{-1}\vek{b}$, and then update the pressures after a small time step $\Delta t$:
\begin{equation}
P_i(t+\Delta t)=P_i(t)+\frac{\partial P_i}{\partial t}\Delta t.
\end{equation}
Based on $V_i=P_i-P_s-I_i^{(c)}R_i^{(c)}$, the fluid content $W_i(t)$ at node $i$ is calculated from the instant value of $P_i(t)$:
\begin{equation}
W_i(t)=C_i\Big(P_i(t)-P_s-R_i^{(c)}\sum_{n(i)}\frac{P_{n(i)}(t)-P_i(t)}{R_{n(i),i}}\Big).
\end{equation}
The simulation proceeds indefinitely toward all $P_i=P_p$ and $W_i^{(\text{max})}=C_i(P_p-P_s)$ which is the maximum fluid content. Theoretically, the process can take an exponentially long time. The average of $W_i(t)$ can be fitted to a function $a\! -\! B\exp(-t/\tau)$ where $a,B,\tau>0$. The time constant $\tau$ is dependent on hydraulic traits, with a unit determined by the product of capacitance and resistance (see discussions in Section \ref{sec:design}). The relative fluid content, $W_i(t)/W_i^{(\text{max})}$, is used in this work to represent the time-varying fluid content distribution in the network.

\subsection{\label{sec:3d} Numerical simulation of the mechanical network and evaluation of the 3D shape}

The simulation methods of the mechanical network, which captures the deformation process (differential expansion) of the simulated surface in three dimensions, are based on a tethered mesh surface discretization model (similar to a spring system) developed for the modeling of deformable membranes and spherical shells \cite{Seung1988, Katifori2010a, Couturier2013}. Figure \ref{network} (b) illustrates a slightly deformed mechanical network, whose buckling is induced by the hierarchy of the flow network in Figure (a) that is overlaid upon it. (Other obtained shapes can be found in the simulation results of Figure \ref{fork} (b).) Also shown are the details of two neighboring faces and their edges and vertices (nodes) in this triangular tessellation of the thin membrane. Each edge behaves like a hookean spring, and the rest length of the bond between nodes $j$ and $k$ (whose positions are $\vek{r}_j$ and $\vek{r}_k$ in 3D space) is $\rho_{jk}$. The elastic stretching energy of the bond is thus $\epsilon_{jk}(\lvert\vek{r}_j-\vek{r}_k\rvert-\rho_{jk})^2/2$, where $\epsilon_{jk}$ is the discretized stretching modulus (spring constant). The elastic surface bending energy, on the other hand, is determined by the angle $\theta_{\alpha\beta}$ made by the normals of neighboring faces $\alpha$ and $\beta$ and is calculated from their respective unit normal vectors $\vek{n}_{\alpha}$ and $\vek{n}_{\beta}$ as $\kappa_{\alpha\beta}(1-\vek{n}_{\alpha}\cdot\vek{n}_{\beta})$ where $\kappa_{\alpha\beta}$ is the discretized bending modulus \cite{Kantor1987, Gompper1996}. This definition is equivalent to an angular energy $\kappa_{\alpha\beta}(1-\cos(\theta_{\alpha\beta}-\theta_{\alpha\beta}^0))$ (which is approximately $\kappa_{\alpha\beta}(\theta_{\alpha\beta}-\theta_{\alpha\beta}^0)^2/2$ when $\theta_{\alpha\beta}\approx\theta_{\alpha\beta}^0$) where the equilibrium angle is $\theta_{\alpha\beta}^0=0$. The total elastic energy of the mechanical network, which is the sum of stretching and bending energies of each bond and between each pair of faces respectively, is defined as:
\begin{equation}
\label{energy}
E=\frac{1}{2}\sum_{\langle jk\rangle}\epsilon_{jk}(\lvert\vek{r}_j-\vek{r}_k\rvert-\rho_{jk})^2+\sum_{\langle\alpha\beta\rangle}\kappa_{\alpha\beta}(1-\vek{n}_{\alpha}\cdot\vek{n}_{\beta})
\end{equation}
where the summations are over adjacent nodes and faces. 

At simulation time $t=0$, the initial surface shape has all bonds relaxed at their rest lengths, which are $\rho_{jk}^0$ at $t=0$. As the simulation proceeds, the rest lengths $\rho_{jk}$ enlarge with time according to the associated fluid contents at the two ending nodes of bonds (e.g.\ $W_j(t)$ and $W_k(t)$). We apply a linear relationship between $\rho_{jk}$ and the average of relative fluid contents at nodes $j$ and $k$, and have:
\begin{equation}
\label{length}
\rho_{jk}(t)=\rho_{jk}^0+\frac{m_{jk}}{2}\Big(\frac{W_j(t)}{W_j^{(\text{max})}}+\frac{W_k(t)}{W_k^{(\text{max})}}\Big)
\end{equation}
where $m_{jk}>0$ so that the surface expands locally as fluid accumulates. Given the fluid content distribution generated from the flow network at time $t$, the total elastic energy $E$ in Equation \eqref{energy} is provided with a particular distribution of $\rho_{jk}(t)$, leading to a change of the shape of the mechanical network which presumably adjusts according to the new rest lengths at timescales much faster than the ones governing the fluid flow. The positions of nodes are adjusted in a quasi-static way in order to minimize $E$, and the system is assumed to have reached mechanical equilibrium before the next fluid flow simulation step at $t+\Delta t$. Note that the shape change is assumed to have no effect on the properties of the flow network. This uni-directional influence is summarized in Figure \ref{network} (c).

\subsection{\label{sec:design} Specifics of network design and architecture}

In this study, we use the same regular triangular lattice for both fluid channels (flow network, see also Reference \onlinecite{Katifori2010}) and surface mesh discretization (mechanical network). The boundary of the network system has a hexagonal shape which is symmetric about its midline (the horizontal long axis at the center in Figure \ref{network}) but non-equilateral. The total number of nodes is 949 including 37 nodes along the midline, and the number of edges is 2740. In the flow network, the same constant hydraulic capacitance $C_i=C$ and resistance to capacitor $R_i^{(c)}=R_c$ are assigned to all the nodes, so that $W_i^{(\text{max})}=C(P_p-P_s)=W_{\text{max}}$ is also a constant. Note that $R_{n(i),i}$ of each vein segment can be non-uniform. In the mechanical network, uniform, constant discretized stretching ($\epsilon_{jk}=\epsilon$) and bending moduli ($\kappa_{\alpha\beta}=\kappa$) are used throughout the whole network. The initial bond rest length $\rho_{jk}^0=\rho_0$ is also identical for all the bonds. We choose a positive constant parameter $m_{jk}=m$ with a length dimension in Equation \eqref{length} for the evaluation of rest length change. The surface shape is initially planar and undeformed at the beginning of a simulation, when all the bonds have the same length and triangular faces in the mesh grid are equilateral. The surface expands differentially as the bond lengths extend quasi-statically according to Equation \eqref{length} (induced by fluid irrigation and storage) in the simulation. When the surface is fully irrigated and all capacitors approximately contain the same amount of fluid $W_{\text{max}}$, the rest lengths of all the bonds are once again identical ($\rho_{jk}=\rho_0+m$), and the surface resumes a planar shape with larger bond length and overall area. In this work, we select $\rho_0=1$, which is used as the basic length unit, as well as $m=1$, so that the rest length increases from 1 to 2 from the beginning to the end of the simulation. We also choose $R_c=1$ and $C=2$ for the simulations, and the basic time unit in this work is hence $CR_c/2$.

The effective thickness $h$ of the elastic membrane and regular triangular network can be calculated by using the relationships between discretized moduli and Young's modulus $Y$ as well as Poisson ratio $\nu$: $\epsilon=\sqrt{3}hY/2$ and $\kappa=h^3 Y/[6\sqrt{3}(1-\nu^2)]$, where $\nu=1/3$ for this discretization \cite{Katifori2010a, Kot2015}. We thus obtain $h=3\sqrt{(1-\nu^2)\kappa/\epsilon}=2\sqrt{2\kappa/\epsilon}$, which is independent of bond lengths or network size and is only dependent on the ratio of discretized moduli. Using $\sqrt{\kappa/\epsilon}=0.306$ (in the length unit $\rho_0=1$) in this modeling, we estimate the effective thickness to be $h=0.8655$, which is thus always smaller than bond lengths and so the system is reasonably a good model for thin membranes.

\begin{figure}[hbt!]
\centering
\includegraphics[width=\textwidth]{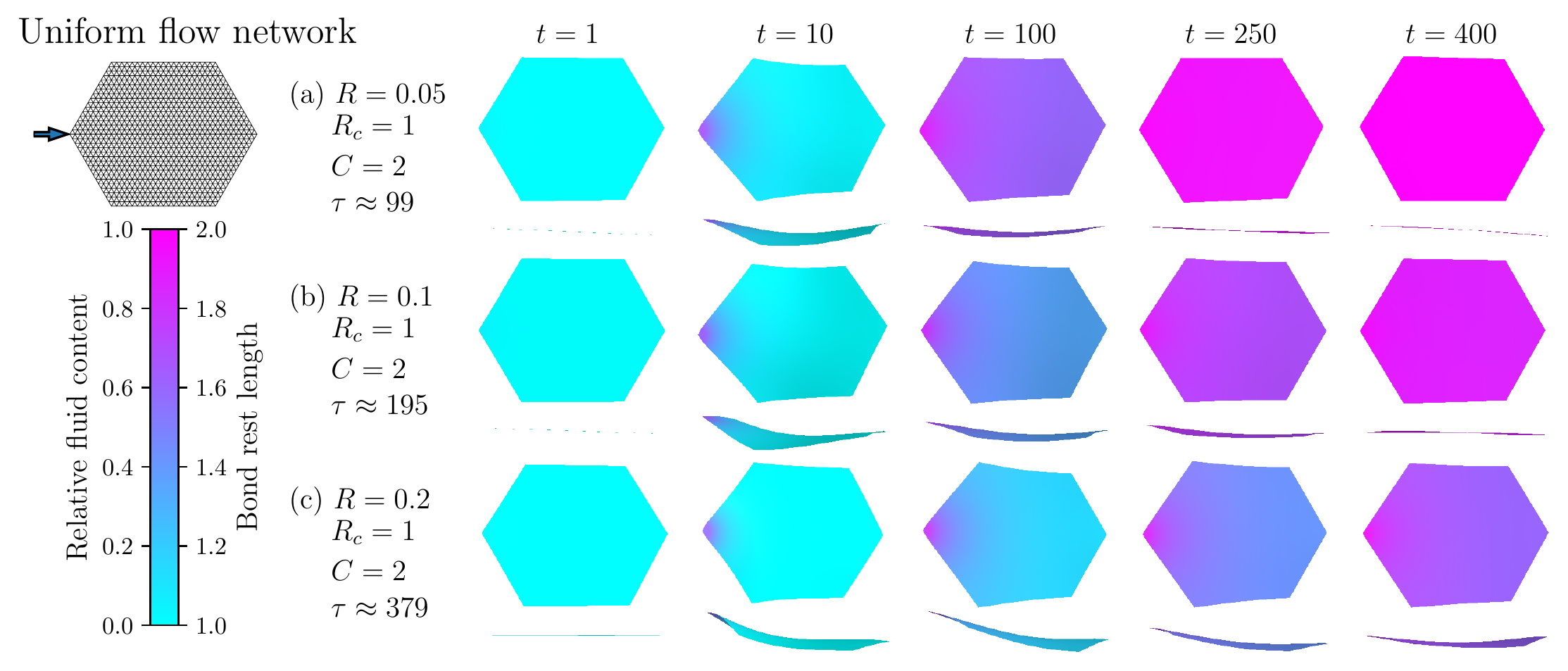}
\caption{Numerical simulation results of time-dependent shape transformations induced by fluid expansion in a uniform flow network. Different resistance parameters are used for (a)--(c) to generate the 3D shapes, whose side views are shown below front views to illustrate possible buckling. With larger resistance, the fluid expansion becomes slower (indicated by larger time constant $\tau$) and the buckling lingers for a longer time.}
\label{uniform}
\end{figure}

In the next section we present the numerical simulation results of several shape transformations, all with the same mechanical design but different hierarchies of hydraulic resistance $R_{n(i),i}$ in the flow network. For the uniform flow network designs in Figure \ref{uniform}, all $R_{n(i),i}=R$ are identical. Different $R$ values are applied along with $R_c=1$ and $C=2$ to study the effect of edge resistance on the transformation dynamics for both fluid spreading and surface expansion. The hierarchical network designs in Figures \ref{fork}--\ref{loop} are based on the parameters of the last uniform network, in which $R=0.2$ is used for the assignment of $R_{n(i),i}$ in a hierarchy. We suppose the cost of each vein segment to be directly proportional to $1/R_{n(i),i}^{\gamma}$ where we choose $\gamma=0.5$ \cite{Katifori2010}, and we keep the total cost of each hierarchy to be the same as that of the uniform network with $R=0.2$, which means $\sum 1/\sqrt{R_{n(i),i}}$ over all the veins is conserved. In each hierarchy, a small resistance value ($R_{\text{maj}}$) and a large value ($R_{\text{min}}$) are used for major and minor veins respectively, which are changed for different designs but always have $R_{\text{maj}}=R_{\text{min}}/1000$. Major veins are thicker in a real-life material and are highlighted with greater thickness in the diagrams. Three categories of hierarchies are designed, including \lq\lq fork-like\rq\rq\ (Figure \ref{fork}, where major vein branches are mostly parallel to the midline), \lq\lq leaf-like\rq\rq\ (Figure \ref{leaf}, where branches extend from the midline), and loopy ones (Figure \ref{loop}, where major veins form loops). 

In each simulation, the fluid always enters the network from the leftmost node on the midline, which is connected to a fluid source across a small resistance at $t=0$. With simulation time step $\Delta t=0.01$, at each integer time point ($t=1,2,\dots$ until $t=500$ for uniform networks and $t=100$ for hierarchies), the instant distribution of relative fluid content is calculated (which is used to get the time constant $\tau$). We then find the shape with minimum mechanical energy $E$ in Equation \eqref{energy}, initializing the optimization  with either a planar shape (at the first time step) or a deformed shape (the optimized shape at a previous time point at subsequent time steps). To quantify the resulting shape, the Gaussian curvature at each node is estimated using an angular deficit method derived from Gauss-Bonnet theorem \cite{Hartig2021}. We show in Figures \ref{uniform}--\ref{loop} the 3D visualizations of obtained shapes, which also illustrate time-dependent fluid distributions in colors, both for uniform networks (which, for our parameters, stay nearly flat) and for various hierarchical structures at certain time points. The 3D renderings are generated from the Mayavi package of Python programming language \cite{Ramachandran2011}, and do not reflect the actual thickness or overall size of the membrane which enlarges gradually, but are plotted to display the surface deformation and buckling. Correlation analyses using linear regression are performed for the fluid distributions and Gaussian curvature arrangements of the resulting shapes of each hierarchical design from $t=1$ to 50. Examples of the analysis performed on the loopy hierarchy are given in Figure \ref{fit}, which also includes 2D plots of relative fluid distribution and Gaussian curvature whose values at each node are linearly interpolated on the surface. Summaries of linear regression results for all hierarchies, as well as more examples of such 2D plots, can be found in Figures \ref{linfork}--\ref{linloop}.

\section{\label{sec:results} Results}
\subsection{\label{sec:uniform} Uniform flow networks}

For a uniform flow network embedded in a surface, such as those shown in Figure \ref{uniform}, the fluid flow resembles uniform diffusion, with a speed of spreading affected by the hydraulic resistances (and capacitance). The rate of fluid spreading and the resulting shape expansion is characterized by a time constant $\tau$ estimated from the temporal changes of average fluid content over the surface. In each simulation series, the membrane keeps a relatively flat shape at the beginning, and then gradually buckles out of the plane, making a slightly curved saddle shape. The curvature of the shape first grows and then diminishes slowly with time as the fluid diffuses spatially, and ultimately after a long period of time, the whole surface is almost uniformly hydrated and the planar shape re-emerges. Apart from estimating $\tau$ from the exponential fitting of average fluid content (see Section \ref{sec:flow}), one can visually compare the expansion speeds of the different simulations by observing both the time and space dependent relative fluid content distributions, which are linearly related to local bond rest lengths and measured by the color bar, and the shape transformations represented by a front view and a side view featuring out-of-plane buckling.

In the flow network, an increase of any of the hydraulic traits $R$, $R_c$ and $C$ will slow down the process, increase the time constant and extend the period during which the surface buckles. This happens as a large resistance hinders the fluid movement and  large capacitors require more time to fill. The effect of a large resistance $R$ of each vein segment (edge), which increases $\tau$ almost twofold when doubled, is directly visible in Figure \ref{uniform}. A selection of larger $R_c$ or capacitance $C$ will be accompanied by a change in the time scale of the simulations, effectively extending $\tau$ and sustaining the buckling deformations for a prolonged period. Based on the selection of large hydraulic parameters in Figure (c), we design different flow network hierarchies to generate various, more pronounced shape changes of the surface.

\begin{figure}[hbt!]
\centering
\includegraphics[width=\textwidth]{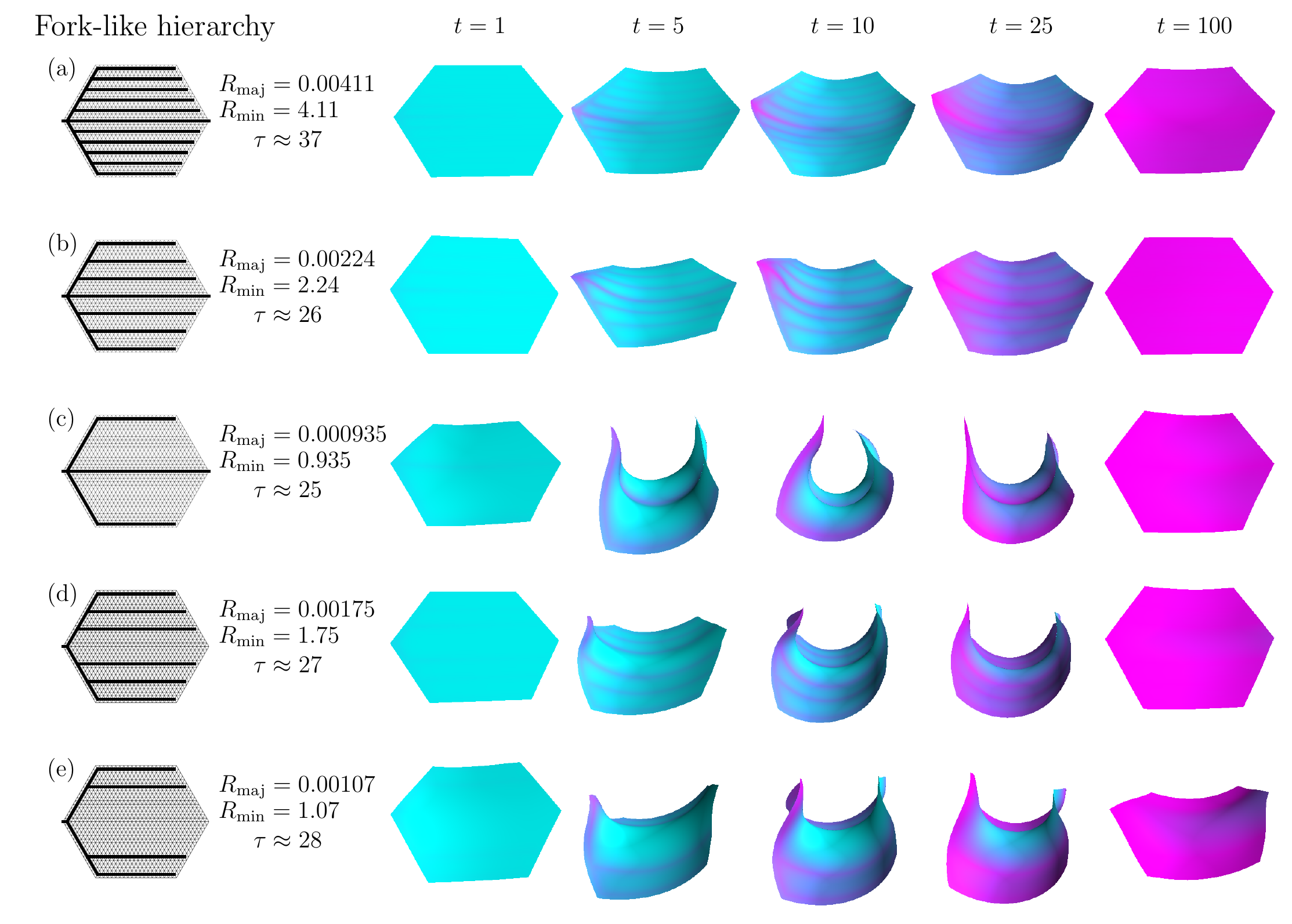}
\caption{Dynamic simulation results of shape transformations induced by flow networks with \lq\lq fork-like\rq\rq\ hierarchies, whose hydraulic resistances of major and minor veins are chosen (in addition to $R_c=1$ and $C=2$) so that the total cost of network is conserved. The 3D visualizations illustrate the instant fluid content distribution (whose average spreading speed is characterized by time constant $\tau$) according to the same color bar of Figure \ref{uniform}.}
\label{fork}
\end{figure}

\subsection{Hierarchical flow networks}

\begin{figure}[hbt!]
\centering
\includegraphics[width=\textwidth]{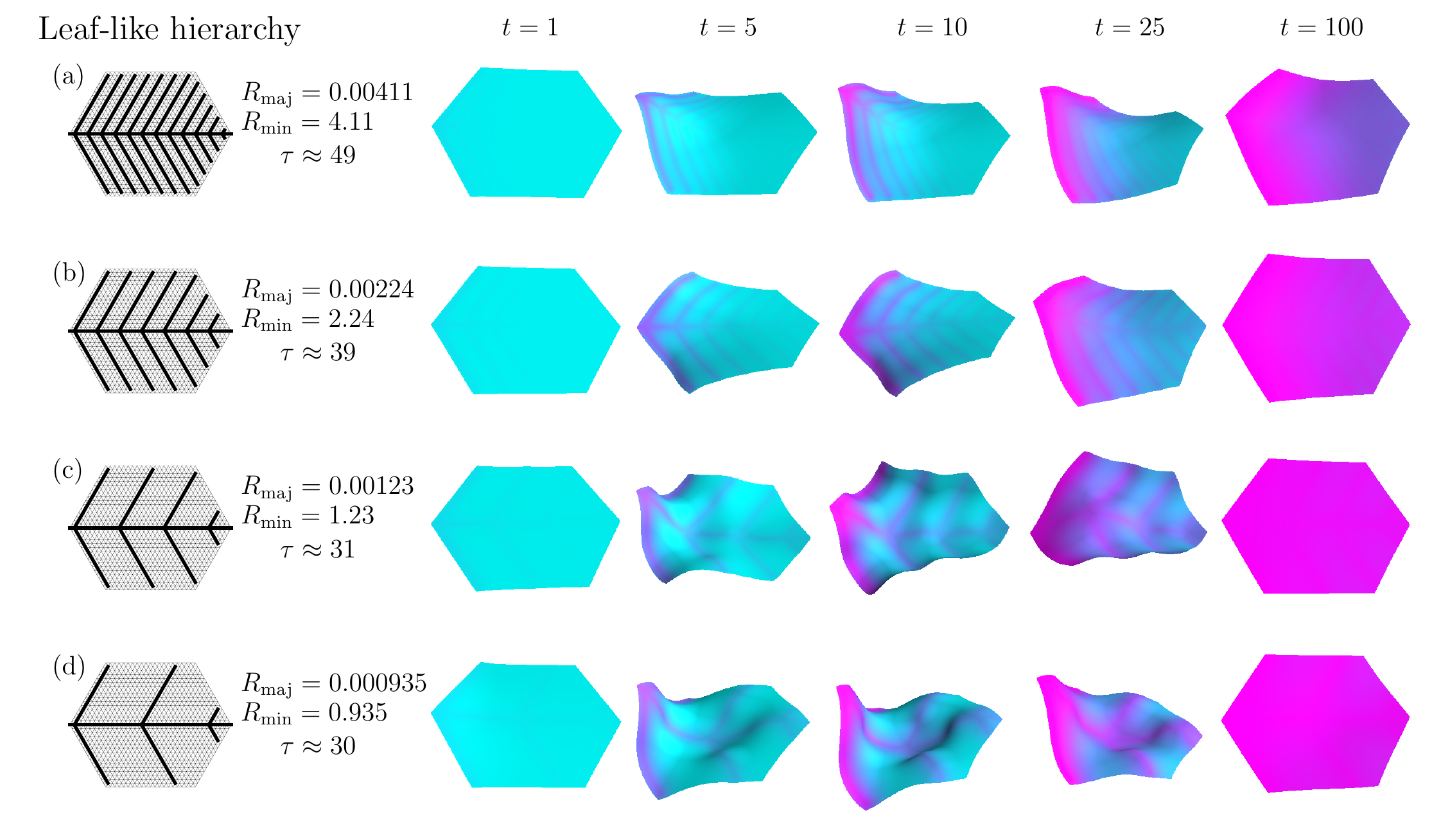}
\caption{Dynamic simulation results of shape transformations induced by flow networks with \lq\lq leaf-like\rq\rq\ hierarchies. See the caption of Figure \ref{fork} for detailed information.}
\label{leaf}
\end{figure}

\begin{figure}[hbt!]
\centering
\includegraphics[width=\textwidth]{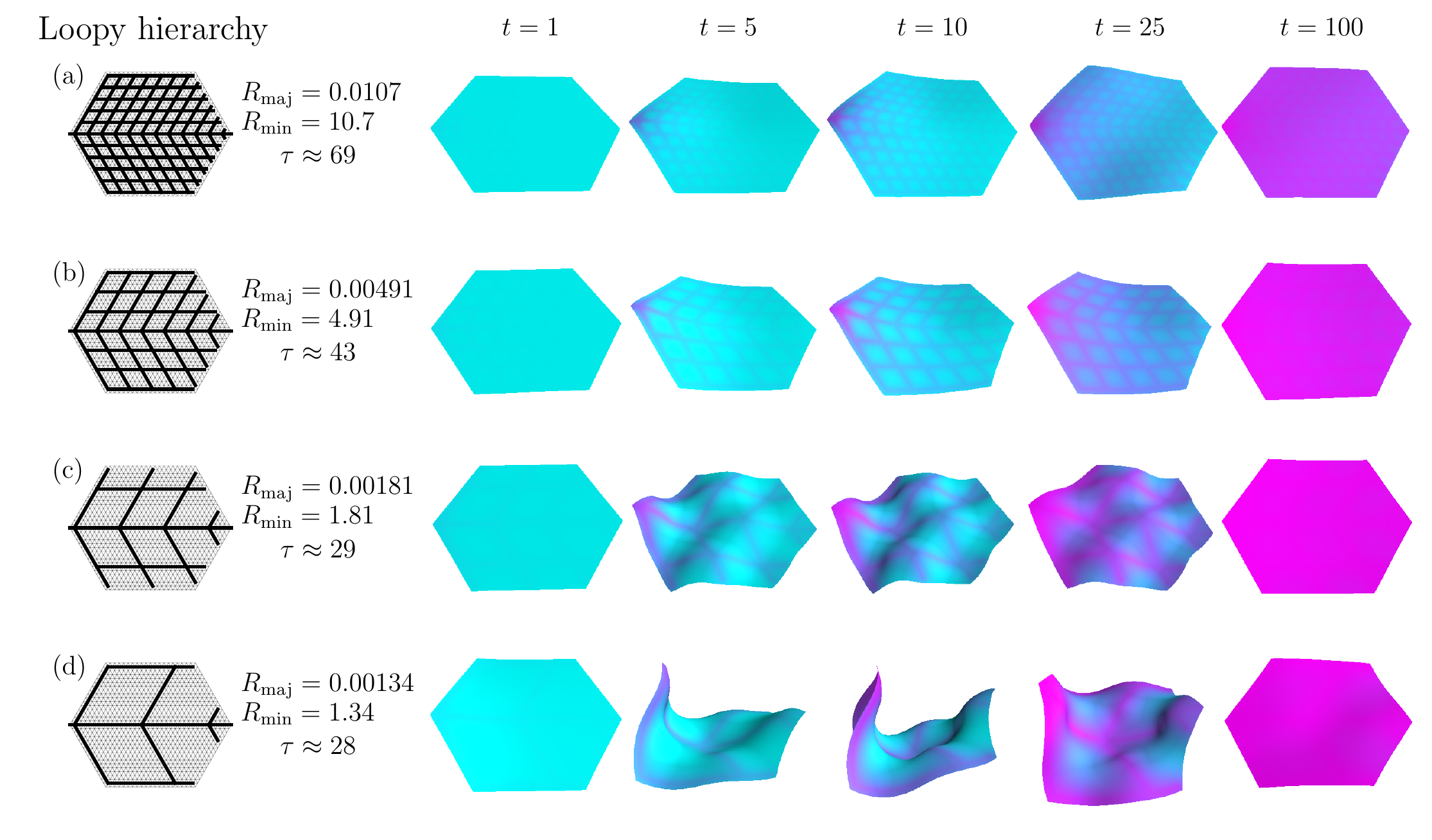}
\caption{Dynamic simulation results of shape transformations induced by flow networks with loopy hierarchies. See the caption of Figure \ref{fork} for detailed information.}
\label{loop}
\end{figure}

The hierarchical network designs, including fork-like, leaf-like and loopy hierarchies, have the same hydraulic parameters used in Figure \ref{uniform} (c) except for $R=0.2$, which is used to derive the major and minor vein resistances for which the total cost of all vein segments is conserved. The high efficiency of fluid transport in major veins with much lower resistance than minor veins is apparent: in Figures \ref{fork}, \ref{leaf} and \ref{loop}, all simulated surfaces starting with an even distribution of fluid (zero everywhere) experience an uneven irrigation process, first through major veins in a fast motion, quickly filling capacitors along these veins, and then through minor veins, gradually spreading into areas farther away in a diffusion-like manner. Correspondingly, the shape of a surface experiences large-scale deformations and buckling induced by differential expansion of the surface because of the non-uniform fluid distribution in the intermediate stages of the fluid flow simulation process. Similar to simulation results of uniform flow networks, after an exponentially long time, the eventually uniform distribution of fluid on the thin surface (nearly fully hydrated everywhere) produces a nearly flat configuration, similar to the starting shape. The areas proximal to fluid source (leftmost node) start to saturate with fluid and deform earlier, and also begin to flatten out earlier than areas on the distal side. The pace of hydration and length of time it takes to expand all over the surface can be characterized by the time constant $\tau$ labeled for each hierarchy.

\begin{figure}[hbt!]
\centering
\includegraphics[width=0.6\textwidth]{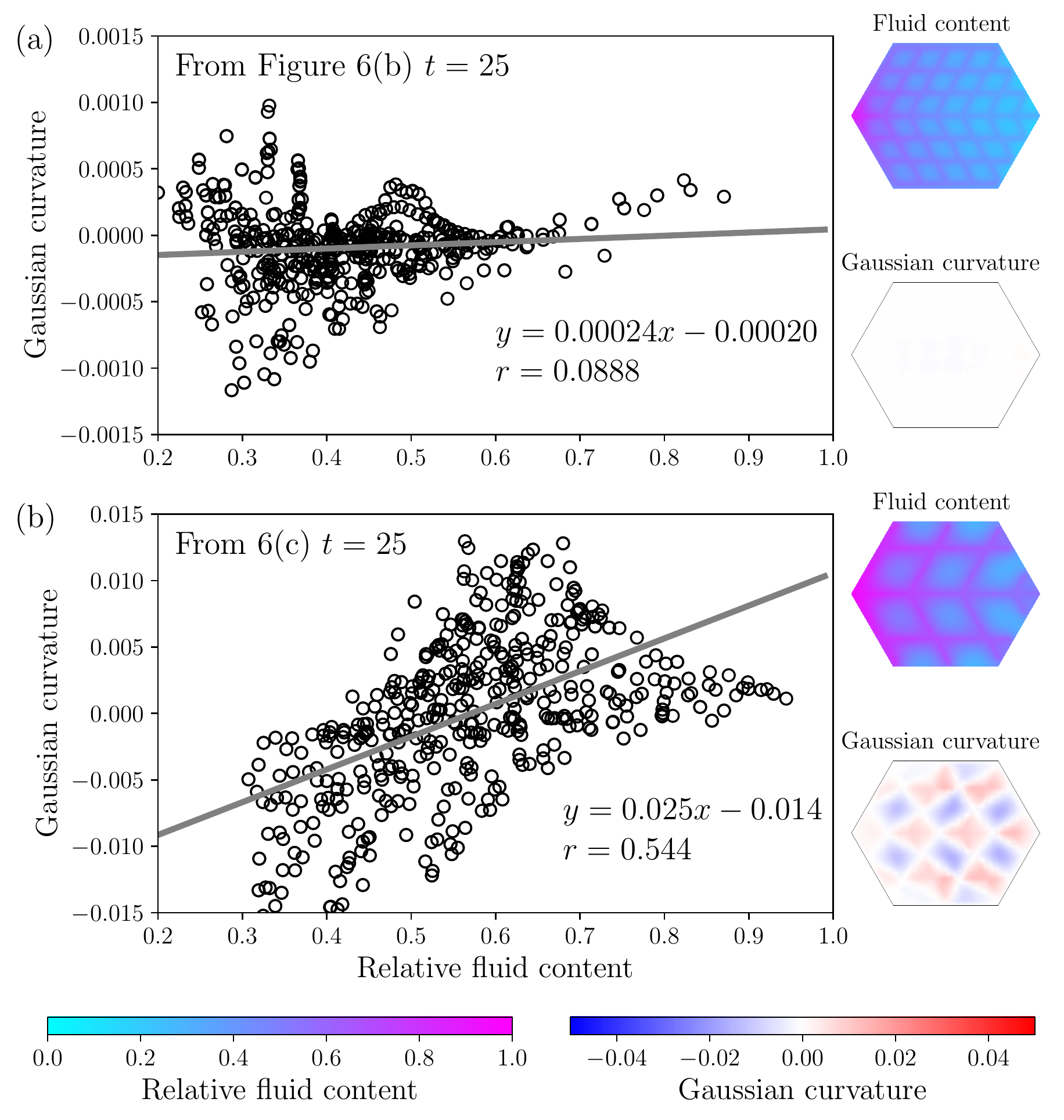}
\caption{Examples of correlation between local relative fluid content and Gaussian curvature over the surface obtained by linear regression at time $t=25$, for the two loopy hierarchical designs from Figure \ref{loop} (b) and (c), respectively. In this figure, plot (a) illustrates a small slope and low degree of correlation (small $r$ value), while plot (b) shows a large slope and high degree of correlation. On the right side of each correlation plot, the 2D plots illustrate both the instant distribution of relative fluid content and that of the estimated Gaussian curvature of the obtained shape.}
\label{fit}
\end{figure}

Two relationships between structure and deformation are easily discernible among all the numerical simulation results of distinct hierarchical flow networks. First, a denser arrangement of major veins (with narrower gaps between major vein branches) leads to a smaller overall curvature change and moderately deformed shapes with less prominent buckling in our dynamic simulation process than a sparser major vein arrangement. Dense arrangements also bring about longer time constants $\tau$, while sparse ones are generally more efficient for fluid movement over the surface on average. Secondly, at a particular time point, areas on the surface with relatively more fluid content tend to form concave or convex shapes sustaining a positive Gaussian curvature, whereas areas with relatively small fluid amount are more likely to form saddle shapes with a negative Gaussian curvature. These correlations increase with the extent of deformation and decline at later time stages as the fluid saturates the capacitors and the shape becomes flat, and are demonstrated with detailed calculation and analyses in Figures \ref{fit}--\ref{linloop}. Both relationships between structure and deformation are most noticeable in the fork-like hierarchy results in Figure \ref{fork}, where the dense major vein connections in (a) and (b) give rise to bowl-shaped deformations with minimal out-of-plane buckling. The magnitude of curvature variation and extent of buckling dramatically grow when major vein branches are spatially more distant from each other such as in (c). There the midline is formed by major veins (with higher instant fluid content) and goes through large positive curvatures before straightening out, while the areas on the two sides of the midline, which are filled with minor veins (with lower fluid content), go through saddle shapes with large negative curvatures. When the midline is formed by minor veins surrounded by parallel major vein branches such as in (d) and (e) where the midline holds less fluid than its surroundings, the resulting saddle-shaped surfaces illustrate negative curvatures around the center and midline. Thus, the hierarchies drastically alter the outcome of specific shapes and curvatures according to the aforementioned observed correlations.

Similar to the fork-like hierarchies, the leaf-like and loopy hierarchies also generate a diverse set of dynamically changing shapes. In Figure \ref{leaf} (a), showing the most dense major vein arrangement, the leaf-like hierarchy leads to overall saddle shapes (unlike the bowl shapes in Figure \ref{fork}). With a smaller major vein density, shown in Figure \ref{leaf} (b), the saddle shapes occur at later times (around $t=25$) after the earlier generation of distorted shapes (around $t=5$ and 10), in which the area near fluid source is twisted at an angle with area far from the source. The transition between the twisted and saddle shapes appears to be discontinuous, and the same discontinuity also exists in the shape change between $t=10$ and 25 in Figure \ref{leaf} (c). The shape transformations in (c) and (d) show the same effects of correlations as summarized above, with more pronounced deformations. 

Last, we study loopy hierarchies by adding major veins parallel to the midline on to leaf-like structures, similar to superimposing the fork-like and leaf-like structures. With dense major veins, the formation of loops hugely reduces curvature development and suppresses large-scale buckling in Figure \ref{loop} (a) and (b). With smaller density (larger gaps) of major veins, the loop formation in Figure \ref{loop} (c) gives rise to fascinating shapes with alternations of concave (or convex) and saddle points on the surface (alternating positive and negative Gaussian curvatures, see Figures \ref{fit} and \ref{linloop}). The loops in Figure (d) appear to promote the development of curvature, bringing about more curved shapes.

\subsection{Correlation analysis of shape and fluid distribution}
The correlation between local deformation and fluid content can be quantified through linear regression of the Gaussian curvature versus relative fluid content, for the optimized surface shapes in Figures \ref{fork}--\ref{loop} at each simulation time point. Figure \ref{fit} specifies this type of analysis for two deformed shapes obtained at time $t=25$ from the loopy hierarchies in Figure \ref{loop} (b) and (c). On the right, the instant distributions of relative fluid amount and Gaussian curvature are plotted in two-dimensional hexagons for the two loopy designs, one with minimal buckling (caused by dense major veins) and the other showing a spatial alternation of positive and negative Gaussian curvatures. In each design (except for peripheral nodes, whose curvatures are assigned zero), the Gaussian curvature at each node is then plotted against its local relative fluid content. The scatter plot including data from all these nodes is used to calculate the least squares linear fitting represented by the solid straight line, and the linear equation and correlation coefficient $r$. Though both correlations are positive indicating that a more positive Gaussian curvature is more likely to emerge from higher fluid content, the architecture with dense major veins shows a weak correlation (small $r$) and the architecture with sparse major veins tend to maintain a strong correlation (large $r$). The magnitude of slope, on the other hand, serves as a measure of shape change, with large slopes indicating prominent deformation. Thus both observations mentioned above about the effects of major vein density are confirmed by this analysis.

\begin{figure}[hbt!]
\centering
\includegraphics[width=0.6\textwidth]{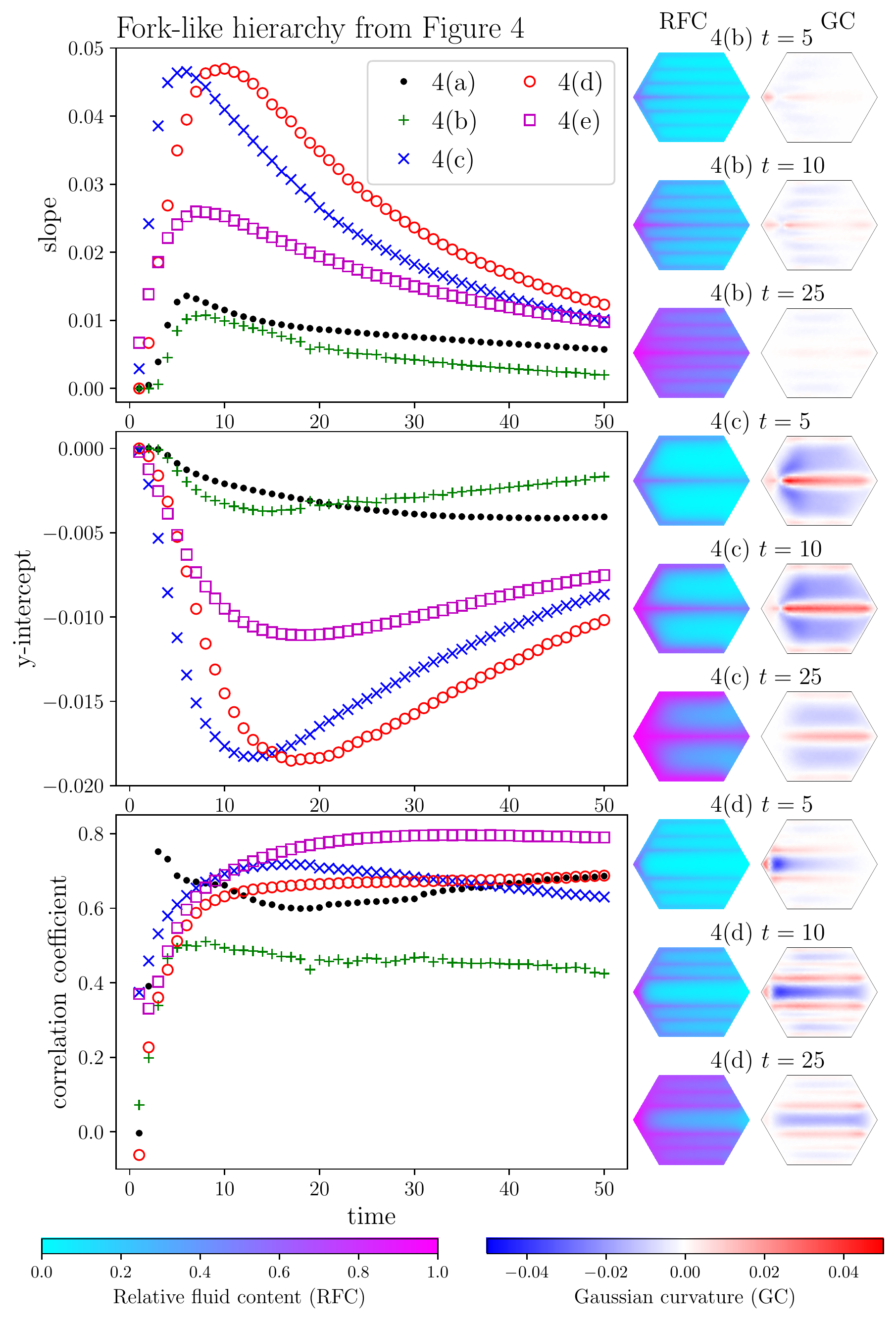}
\caption{The correlation between local relative fluid content and Gaussian curvature over the surface, as represented by the linear regression results for each \lq\lq fork-like\rq\rq\ hierarchy at each integer time point from $t=1$ to 50. On the right side, 2D plots of fluid content (RFC) and Gaussian curvature (GC) distributions are shown for several hierarchies at certain time points (corresponding to some shapes in Figure \ref{fork}). The symbols 4(a)--4(e) represent vein patterns with the same labels in that figure.}
\label{linfork}
\end{figure}

\begin{figure}[hbt!]
\centering
\includegraphics[width=0.6\textwidth]{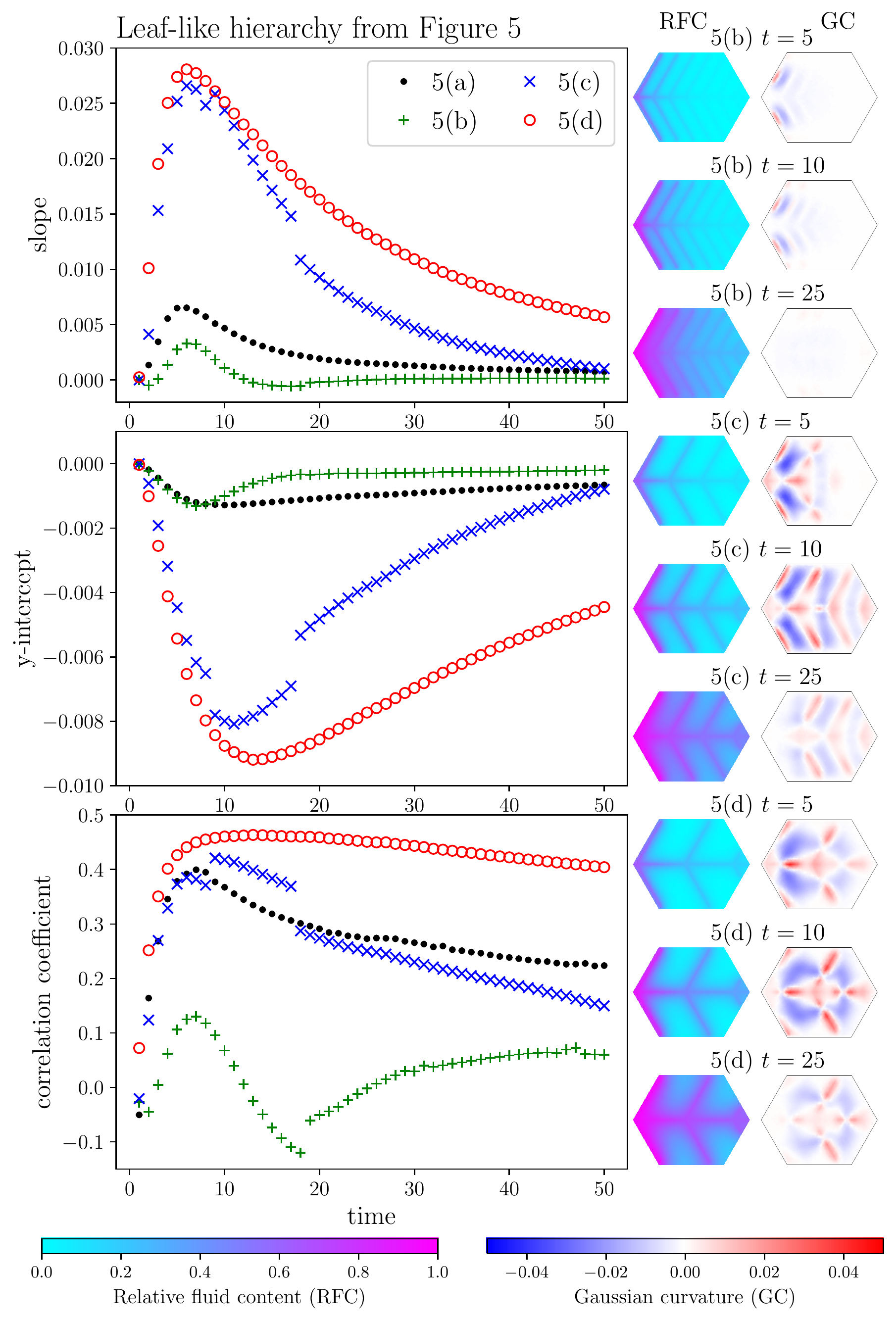}
\caption{Linear regression results for each \lq\lq leaf-like\rq\rq\ hierarchy, with 2D plots corresponding to some shapes in Figure \ref{leaf}. The symbols 5(a)--5(d) represent vein patterns with the same labels in that figure. See the caption of Figure \ref{linfork} for more information.}
\label{linleaf}
\end{figure}

\begin{figure}[hbt!]
\centering
\includegraphics[width=0.6\textwidth]{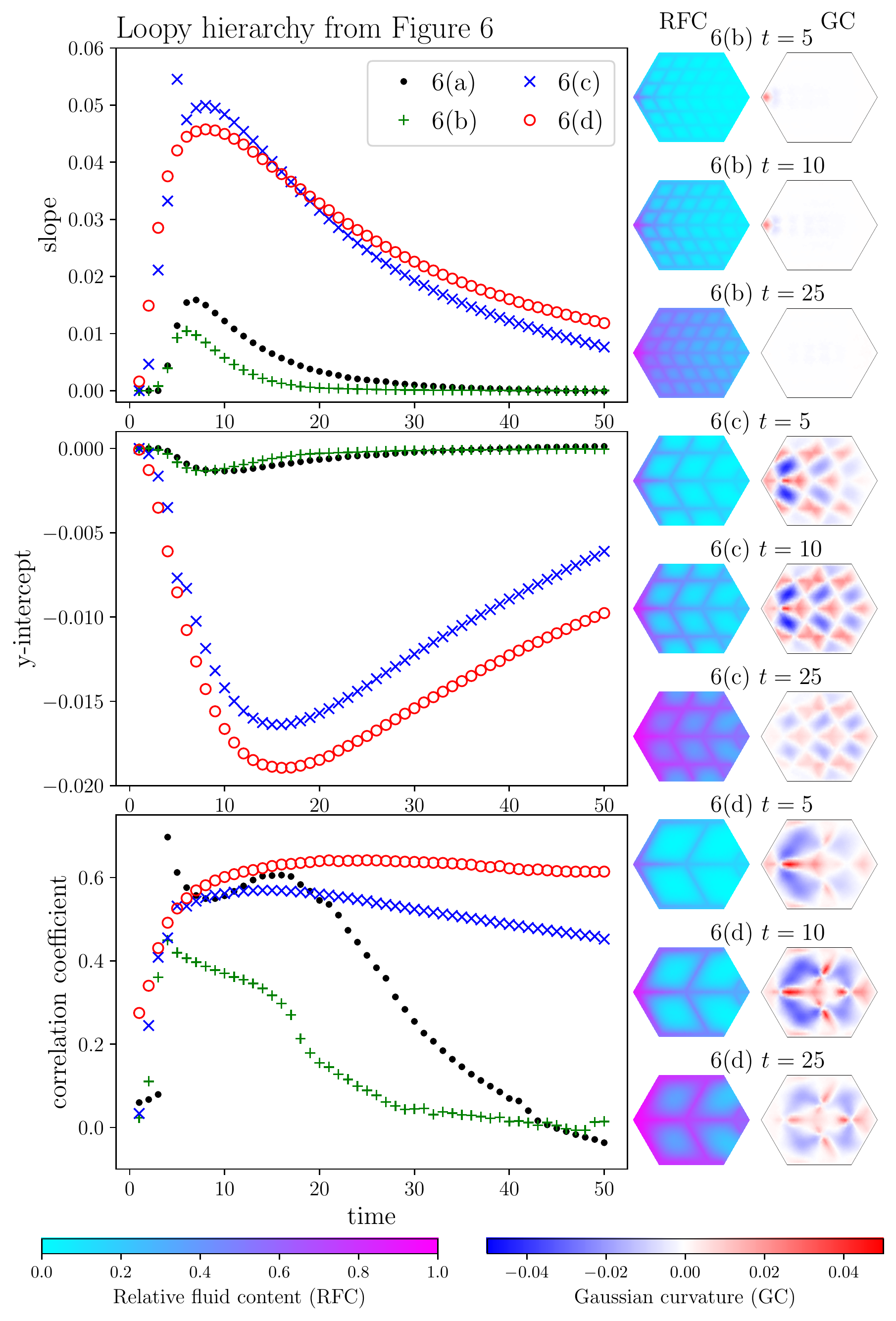}
\caption{Linear regression results for each loopy hierarchy, with 2D plots corresponding to some shapes in Figure \ref{loop}. The symbols 6(a)--6(d) represent vein patterns with the same labels in that figure. See the caption of Figure \ref{linfork} for more information.}
\label{linloop}
\end{figure}

Correlation analyses carried out by using the same method as in Figure \ref{fit} (containing linear fit slope, y-intercept and correlation coefficient, all of which vary with time) and applied to fork-like, leaf-like and loopy hierarchy results (optimized surface shapes) from time $t=1$ to 50, are summarized in Figures \ref{linfork}, \ref{linleaf} and \ref{linloop}, respectively. Similar 2D hexagonal plots to those of relative fluid content and Gaussian curvature in Figure \ref{fit} are generated for several selected shapes found in Figures \ref{fork}--\ref{loop} (b), (c) and (d). They can be located on the summary plots of the linear regression results. In general, the resulting slopes are positive and y-intercepts are negative. The correlation coefficients (as well as both slopes and y-intercepts) are generally larger in magnitude for designs with less dense major vein arrangements (wider gaps between major vein branches). These findings generalize the observations from Figure \ref{fit} about positive correlations and their strengths (and also intensities of deformation) in different hierarchies to basically all obtained shapes at all times, with sparser major vein connections giving rise to stronger correlations (and larger deformations). The dynamic variations of linear fit results follow comparable trends with time for all hydraulic hierarchical categories. The correlation sharply increases initially starting with empty capacitors and undeformed surface, rapidly growing as represented by the correlation coefficient along with the intensive shape transformation and curvature development (in terms of slope, with y-intercept becoming more negative) that happen simultaneously. After reaching a peak correlation (which is also near the point of maximum deformation and curvature forming of the surface) it finally diminishes in a slow pace, corresponding to the gradual decline of curving and buckling to a uniformly expanded flat shape with full capacitors. In this process, the Gaussian curvature plots clearly exhibit the movement of curvature formation from the near side of fluid source (left side of the surface) to the far side (right side) following the fluid spread and differential expansion. The discontinuous shape transitions in leaf-like hierarchies of Figure \ref{leaf} (b) and (c) are also reflected by the discontinuity in one or more of the linear regression outcomes at around $t=18$ in Figure \ref{linleaf}, for example the breaking of correlation coefficient trend for both (b) and (c) and also the jumping of slope and y-intercept for (c).

\section{\label{sec:disc} Discussion}

Our work makes a unique contribution to the large volume of literature regarding the shape morphing of thin membranes related to hydraulics, by revealing the significant roles of fluid delivery and storage capability. We show that besides helping to withstand drought conditions \cite{Luo2021}, the large hydraulic capacitance in petals can also contribute to the extension of petal buckling and unfolding period (see Section \ref{sec:uniform}), affecting the timescale of opening in many species. Our work, however, was only inspired by biology, and ignores biological complexity, such as the biochemistry that might affect the storage capacity of the cells. In order to investigate the extent to which our findings are relevant for real-life petals, experimentalists can apply novel, advanced measurement approaches on petals, such as time-lapse imaging and video of deformation processes \cite{Portet2022}, and also innovative techniques to extract flow network architecture \cite{Katifori2012, Vasco2014} or to keep track of fluid movement using fluorescent dyes \cite{Katifori2010, Liu2016}.

The existence of hierarchies in vascular flow networks was shown to optimize transport efficiency (with branches \cite{Ronellenfitsch2016}) and be robust to damages and fluctuations (with loops \cite{Ronellenfitsch2019}). Similarly, in mechanical networks, the network rigidity was shown to be strengthened by hierarchical elastic structures \cite{Ronellenfitsch2021}. In this work, we demonstrate another straightforward impact of the topology of venation on thin sheet mechanics, through local fluid absorption, differential expansion and ensuing deformation. With fluid quickly spreading through low-resistance major vein highways (confirmed experimentally in leaves \cite{Zwieniecki2002}), fluid storage elements are rapidly filled up in their vicinity undergoing local growth. Unless there exists a nonlocal feedback control mechanism, the surface cannot remain flat when growing \cite{Al-Mosleh2022}.

The differential expansion due to fluid storage forces the surface to bend into a buckled shape, in which positive Gaussian curvatures build up in relatively liquid-rich areas and negative ones form in liquid-poor regions. The emergence of saddle shapes similar to those in Figures \ref{fork} and \ref{leaf} has been observed in leaves and petals and modeled using differential growth, in addition to rippling patterns near the surface margin due to edge elongation \cite{Liang2009, Liang2011, Sharon2004}. Similar patterns are produced by some of our modeling results (Figures \ref{leaf} (c) and \ref{loop} (c)) with an overall surface undulation (because of the spatial alternations of positive and negative Gaussian curvatures). The undulating shapes in this work, controlled by flows, can be compared to patterns emerging on confined, intrinsically curved thin shells arising from their surface area mismatch with the geometry of the confinement \cite{Aharoni2017, Albarran2018, Tobasco2022}. Hence, our work provides an alternative mechanism for the appearance of nontrivial deformations of a thin shell in a physical or biological system in addition to well established methods. 

These deformation mechanisms, including the one in this study, tend to emphasize the geometric aspects of shape transformations by focusing on surface growth and expansion. However, mechanical factors can be easily introduced to our model, by having stretching and bending moduli in Equation \eqref{energy}, rather than just bond rest length, also increase with the buildup of local fluid content. This would correspond to a differential stiffening, which may accelerate the restoration of flat shape.

Our work focuses on analyzing how the topology and hierarchy of irrigation networks affect the flow-controlled deformation dynamics of the fluid absorbing membrane. The connectivity of our model flow network was selected to represent broad classes of biologically inspired networks. All of the \lq\lq fork-like\rq\rq, \lq\lq leaf-like\rq\rq\ or loopy hierarchies of major veins were readily observed in the micrographs of flower petals of several dicot species \cite{Zhang2017, Roddy2019}. The branching designs resembling leaves (Figure \ref{leaf}) and reticulate ones with loops (Figure \ref{loop}) are actually inspired by the model leaf venations in Reference \onlinecite{McKown2010}. Some network models were developed as inspired by the leaf of a monocotyledon (like wheat \cite{Altus1985}) or dicotyledon (like laurel \cite{Cochard2004}), in which square grids were applied and multiple levels of resistances were assigned to different degrees of veins. In this work, on the contrary, we only apply two resistance values, different by a factor of 1000, to major and minor veins to focus on the fundamental aspects of hierarchy. We also implement a triangle, rather than a square, mesh grid with hexagonal boundary which was also used for leaf models previously \cite{Katifori2010, Katifori2012}. The triangular grid of the fluid flow system can be readily used for the architecture of the mechanical network that models the elastic properties and shape changes of the thin sheet surface, as triangle grid meshes capture both stretching and bending. Our simple model with only very basic components can be easily augmented to incorporate spatiotemporal inhomogeneities into the flow and mechanical networks. For example, there can be more than two levels of vein resistance to capture the tapering and reduction of conductance in major veins from petiole (fluid source) to the margin \cite{McKown2010}. The effect of transpiration can be added by setting internal nodes as flow sinks, and the mechanical properties can be spatiotemporally variable, reflecting the in-plane anisotropy of the thin material (e.g.\ the diversification of petal cell elasticity) or more complicated effects of fluid accumulation.


The numerical findings in this work, especially the local correlations between surface curvature and fluid content in the thin membrane, can also provide helpful insights into how the conformation of a swellable thin material can be fine-tuned. This fine tuning is achieved by adjusting the fluid distribution. Given a target surface configuration, the arrangement of curvatures can be calculated, and then correlated to relative fluid contents through the correspondences found in Figures \ref{fit}--\ref{linloop}. Though the shapes created in this work are dynamical and transient, one can potentially program the surface into a desirable shape with high confidence (based on a linear relationship with strong correlation) by holding the underlying local fluid contents in a steady state through an arresting mechanism. 

Formerly, by manipulating deforming mechanisms like differential growth, engineers have fabricated thin sheets responsive to environmental stimuli, making a large variety of structures including nematic glass cantilevers \cite{Warner2010, Warner2010a} and temperature-sensitive copolymer disks \cite{Kim2012}, which can buckle when being heated or cooled. Expansive computational studies and analytical calculations have also been carried out, in order to design real physical systems that harness differential swelling to transform into intended shapes when needed \cite{Dias2011, Nardinocchi2015}. Our work serves as a complement to this field by involving fluid flow and storage as a controlling factor behind the swelling. A similar actuation principle was recently applied to macroscopic air-filled rubber plates embedded with airway channel networks, which cause shape morphing when being inflated \cite{Siefert2019, Siefert2020}. For this work, notably, hydrogel polymers, which enlarge in size when immersed in water and have been made into numerous conformations in biomimetic 4D printing \cite{Gladman2016}, appear to be an ideal material for the fabrication of deformable thin sheets implanted with a microfluidic channel network, which can possibly test our modeling results. Microchannels embedded in thin films have already been used to simulate plant hydraulics, with one example in silicone PDMS \cite{Noblin2008} and another in hydrogel PHEMA \cite{Wheeler2008}, and also embedded in liquid crystal elastomer sheets to pre-program intrinsic spontaneous curvatures \cite{Xia2016, Aharoni2018}. The combination of microfluidics and micromechanics in a hydrogel-based soft membrane prone to flows can recreate the coupling effects in our network model, whose major veins are explicitly mimicked by the channels and minor veins implicitly approximated by the rest of the surface where fluids diffuse and remain, and then the dynamic deformation pathways can be scrutinized physically. Overall, our study in this work reveals a plethora of possibilities in physics, biology and applied science for future theoretical, computational and experimental investigations to further discover.

\appendix
\section{Experimental method to capture petal fluorescent images in Figure \ref{inspire} (b)}
\emph{Hosta sp.} was collected at the University of Pennsylvania, brought back to the lab, and re-cut under water immediately. Right before the measurement, the flower was cut, flattened, and affixed onto a black background using clear tape. At time 0, the cut-end of the pedicel was submerged in a petri dish with 0.1\% fluorescein solution (fluorescein free acid, Honeywell Fluka, in \SI{15}{\milli\Molar} KCl solution of pH 8). The petri dish was covered with foil paper to prevent any fluorescent noise. Images were taken in a photo light box to eliminate any light source from the lab. In the photo light box, the light was provided by an LED growth light strip with PPFD circa \SI{300}{\micro\mol\cdot{\m}^{-2}{\s}^{-1}}. Images were taken every 120 seconds with a Nikon D3300 camera and a color blocking filter (Filter Orange \#22, Heliopan, Munich) mounted on a tripod stand and connected to a PC with the digiCamControl software.


\begin{acknowledgements}
The authors acknowledge support from NSF-IOS, award 1856587, and thank Shu Yang for helpful discussions.
\end{acknowledgements}

\bibliography{main}

\end{document}